# Prosperity is associated with instability in dynamical networks

Matteo Cavaliere[1,3,#], Sean Sedwards[1,4,#], Corina E. Tarnita[2], Martin A. Nowak[2], Attila Csikász-Nagy[1,*]

[1] The Microsoft Research–University of Trento Centre for Computational and Systems Biology, Povo (Trento) 38123, Italy

[2] Program for Evolutionary Dynamics, Harvard University, Cambridge, MA 02138, USA

[3] Present address: Logic of Genomic Systems Laboratory, Spanish National Biotechnology Centre (CNB-CSIC), Madrid 28049, Spain

[4] Present address: INRIA Rennes – Bretagne Atlantique, Campus universitaire de Beaulieu, 35042 Rennes Cedex, France

[#] These authors contributed equally to this work.

[*] To whom correspondence may be addressed. E-mail: csikasz@cosbi.eu

## Abstract

Social, biological and economic networks grow and decline with occasional fragmentation and re-formation, often explained in terms of external perturbations. We show that these phenomena can be a direct consequence of simple imitation and internal conflicts between 'cooperators' and 'defectors'. We employ a game-theoretic model of dynamic network formation where successful individuals are more likely to be imitated by newcomers who adopt their strategies and copy their social network. We find that, despite using the same mechanism, cooperators promote well-connected highly prosperous networks and defectors cause the network to fragment and lose its prosperity; defectors are unable to maintain the highly connected networks they invade. Once the network is fragmented it can be reconstructed by a new invasion of cooperators, leading to the cycle of formation and fragmentation seen, for example, in bacterial communities and socio-economic networks. In this endless struggle between cooperators and defectors we observe that cooperation leads to prosperity, but prosperity is associated with instability. Cooperation is prosperous when the network has frequent formation and fragmentation.

## 1 Introduction

Networks interpreted as graphs, consisting of nodes linked by edges (Erdős and Rényi, 1960), are used to model a wide variety of natural and artificial systems (Barabasi and Albert, 1999; Boccaletti et al., 2006; Csermely, 2009; Dorogovtsev and Mendes, 2003; Jackson, 2008; Montoya et al., 2006; Newman et al., 2001; Watts and Strogatz, 1998). The evolution and formation of complex networks has been widely investigated (Boccaletti et al., 2006; Dorogovtsev and Mendes, 2003), often with the goal of understanding how certain topologies arise as the result of copying interactions (Davidsen et al., 2002; Jackson and Rogers, 2007; Kleinberg et al., 1999; Krapivsky and Redner, 2005; Kumar et al., 2000; Rozenberg, 1997; Sole et al., 2002; Vazquez et al., 2001). Indeed, imitation is ubiquitous and is often crucial in systems where global knowledge is not feasible (Bandura, 1985). This mechanism can be conceptually divided into two components: the imitation of behaviours (strategies) and the imitation of social networks (connections). For instance, in networks where links denote interpersonal ties, newcomers want to establish links to successful people, imitate their behaviour and connect to their friends (Jackson, 2008; Jackson and Rogers, 2007); in financial networks (Bonabeau, 2004; Schweitzer et al., 2009) the links are business relationships where newly created institutions emulate the successful strategies of other institutions and try to do business with the same partners; (Haldane, 2009b). At a completely different scale, in bacterial communities and multicellular systems, where the links denote physical connections, a successful cell duplicates and its offspring inherit ('imitate') the strategies (genomes) and the connections of its parents. Several studies have shown the general relevance of imitation to the outcome of social dilemmas in well-mixed and structured populations (Hofbauer and Sigmund, 1988; Lieberman et al., 2005; Nowak, 2006b; Nowak and Sigmund, 2004; Ohtsuki et al., 2006; Pacheco et al., 2006; Szabó and Fáth, 2007) and to the dynamics of complex systems and networks (Akerlof and Shiller, 2009; Bonabeau, 2004; Castellano et al., 2009; Haldane, 2009b; Helbing, 2010; Sornette, 2003), but it is an open challenge to understand the role of imitation in the interplay between individual benefits and the overall prosperity and stability of a system (Bascompte, 2009; Haldane, 2009b; Haldane and May, 2011; Jackson, 2008; Schweitzer et al., 2009).

To address this challenge we employ a game theoretical model of dynamic networks where nodes can be cooperators or defectors and newcomers imitate the behaviour (strategies) and the social network (connections) of successful role-models. We show that the recurrent collapses and re-formations that characterize certain natural and manmade systems, often investigated in terms of external perturbations (Albert et al., 2000; Billio et al., 2010; Haldane, 2009b; Montoya et al., 2006; Paperin et al., 2011; Scheffer et al., 2009), can be explained in our model as the consequence of imitation and endogenous conflicts between cooperators and defectors.

Cooperation leads to prosperity, but we observe that prosperity is associated with increased connectivity, which in turn makes the network vulnerable to invasion by defectors and network collapse. Thus, the long-term prosperity and stability of the system are negatively correlated and we find that cooperation is most productive when the system is unstable.

## 2 Methods

### 2.1 Model

We consider a network of $N$ nodes linked by a number of edges which varies over the course of the evolution of the system. Each node in the network adopts one of the two strategies of the *Prisoner's Dilemma* (Hofbauer and Sigmund, 1988; Nowak, 2006a; Nowak and Sigmund, 2004): a *cooperator* pays a *cost c* to provide a *benefit b* to all of its neighbours; *defectors* pay no cost and distribute no benefit. At each step and for each node i, $Payoff_i$ is calculated as the sum of pair-wise interactions with its neighbours[1]. A new node (a *newcomer*) is then added and a randomly chosen existing node is removed from the system.

A node is selected probabilistically from the population to act as *role-model* for the newcomer. The probability of a node *i* to be selected as a role-model is proportional to its *effective payoff* $EP_i = (1+\delta)^{Payoff_i}$, where $\delta \geq 0$ specifies a tuneable intensity of selection (the exponential function affords the model greater flexibility without losing generality (Aviles, 1999; Traulsen et al., 2008)). A newcomer copies its role-model's strategy with probability 1-*u* or mutates to the alternative strategy with probability *u*. The newcomer is then *embedded* into the network: it establishes a connection with each of the role-model's neighbours ('copies' its connections) with probability *q* and directly with the role-model with probability *p*. Thus, with probability $q^k$ a newcomer connects to all *k* neighbours of the role-model. Hence, the parameter *u* controls the chance to imitate the strategy of a role-model, while the parameters *p* and *q* explicitly model the ability to imitate the role-model's social network and are referred to as *embedding parameters* because they control how the newcomer is embedded in the network. Notice that the number of nodes is maintained constant during the evolutionary process. In this respect, our model works along the lines of a Moran process, which describes the evolution of finite resource-limited populations and allow some analytical simplicity (Moran, 1962; Nowak, 2006a). A diagrammatic description of the model is given in Figure 1.

---

[1] E.g., if a cooperator node has *C* cooperator neighbours and *D* defector neighbours, its *Payoff* is *C*(*b*-*c*)-*Dc*. A defector node in the same neighbourhood has *Payoff* = *Cb*.

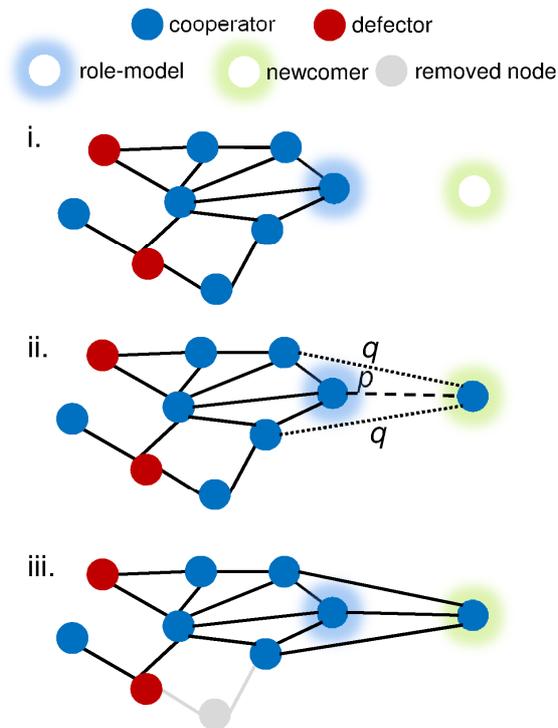

**Fig. 1: Network update mechanism.**
Newcomers imitate the strategy and social network (connections) of successful role-models: (i) A role-model is selected based on its effective payoff. (ii) The newcomer connects to the role-model with probability *p* (dashed line), connects to each of its neighbours with probability *q* (dotted lines) and emulates its strategy with probability *u*. (iii) A randomly selected node and all its connections is removed from the network.

## 2.2 Simulations

Computer simulations and visualisations were performed using custom created software tools written in Java[2]. Simulation runs start from a randomly connected network of $N$ nodes[3] having average connectivity $k = 4$ and proceed with a sequence of $10^8$ steps, as described in Section 2.1. All nodes initially adopt the same strategy and long term statistics are calculated by taking the average of two runs, one starting with all cooperators, the other with all defectors, excluding the first $10^6$ steps. At each step the total effective payoff of a network is calculated as $EP_{tot} = \sum_{i \in \{1...N\}} EP_i$. The probability to choose a node as role model is then $EP_i/EP_{tot}$. Hence, $\delta = 0$ produces a uniformly random choice of node, independent of payoff, while increasing $\delta$ makes it more likely to choose nodes with higher payoffs. We define *prosperity* as $100 \cdot (\sum_{i \in \{1...N\}} Payoff_i) / (N \cdot (N-1) \cdot (b-c))$, i.e. the total payoff of the network as a percentage of the total payoff of a fully-connected network of cooperators.

*Long term* cooperation, connectivity, largest component and prosperity are calculated as the sum of the number of cooperators, average node degree, number of nodes in the largest component and prosperity at each step, respectively, divided by the total number of steps considered.

---

[2] An online companion software tool that reproduces our results can be found at www.dynamicalnetworks.org

[3] Random networks are generated by making any particular link with probability $k/(N-1)$

# 3 Results

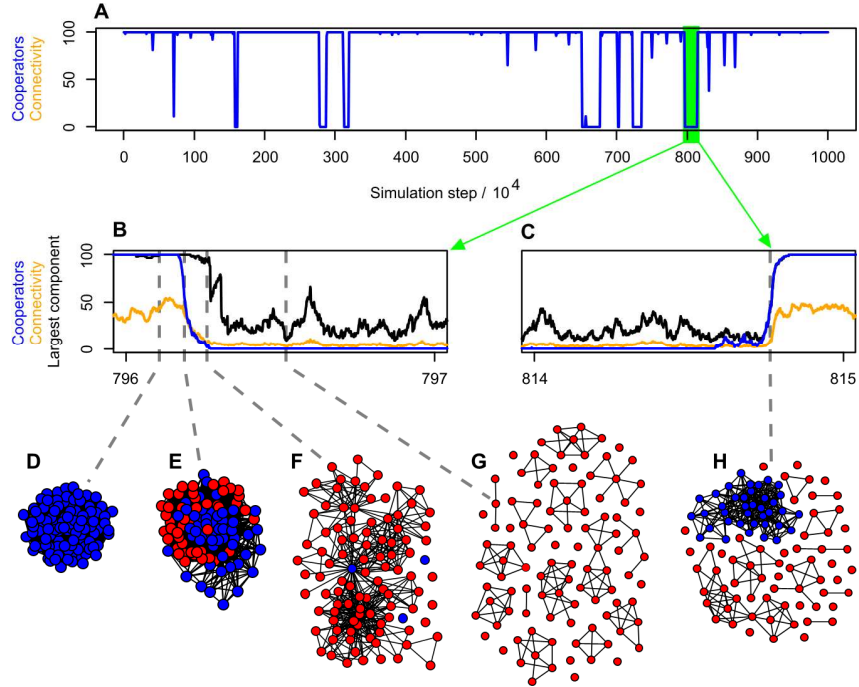

**Fig. 2: Typical simulation run that favours cooperators but features transient invasions of defectors.** A network of $N = 100$ nodes is simulated with parameters $b/c = 3$, $p = 0.6$, $q = 0.85$, $u = 0.0001$ and $\delta = 0.01$. (**A**) Fluctuating abundance of cooperators. (**B**) Transition from all-cooperators to all-defectors accompanied by network fragmentation. (**C**) Transition from all-defectors to all-cooperators showing the synchronous rise in the size of the largest component. (**D-H**) Graphical depiction of networks during the transitions of (B) and (C): (**D**) a highly connected network of cooperators (blue); (**E**) defectors (red) invade the network, causing a reduction in connectivity; (**F**) few cooperators remain and the network is becoming sparsely connected; (**G**) with only defectors present the network disintegrates; (**H**) a single component of cooperators start to reconstruct the network. The end result of this construction process is a network which resembles that of (D).

When mutation is rare, we observe *transitions* between the extreme states consisting of all cooperators and all defectors (Fig. 2). Such transitions are typically associated with changes of network topology. When defectors take over, the network disintegrates, while the dominance of cooperators is associated with more connected networks. The network tends to contain a large, well-connected component as long as cooperators are prevalent, while the network becomes fragmented into many smaller components when defectors prevail. During a transition from cooperation to defection, the network fragments only after defectors have taken over (Fig. 3A). For a transition in the opposite direction, the construction of the network is synchronous with the rise of cooperators (Fig. 3B). We also note that the delay between the spreading of defectors and the network fragmentation is an increasing function of the embedding parameters, while the time for the network to rebuild is a decreasing function of those parameters (Fig. 3). These phenomena are robust for a wide range of parameters and initial conditions, as well as when newcomers are drawn from the existing population and 'remember' some of their previous connections (see Electronic Supplementary Information). Thus, despite the fact that cooperators and defectors are embedded and removed in an identical way, we observe that cooperators promote the formation of well-connected networks and defectors cause the network to fragment.

The way newcomers are embedded into the network influences the topology of the network, which in turn affects the ability of cooperators to resist invasion by defectors and to reconstruct the network once it has been destroyed.

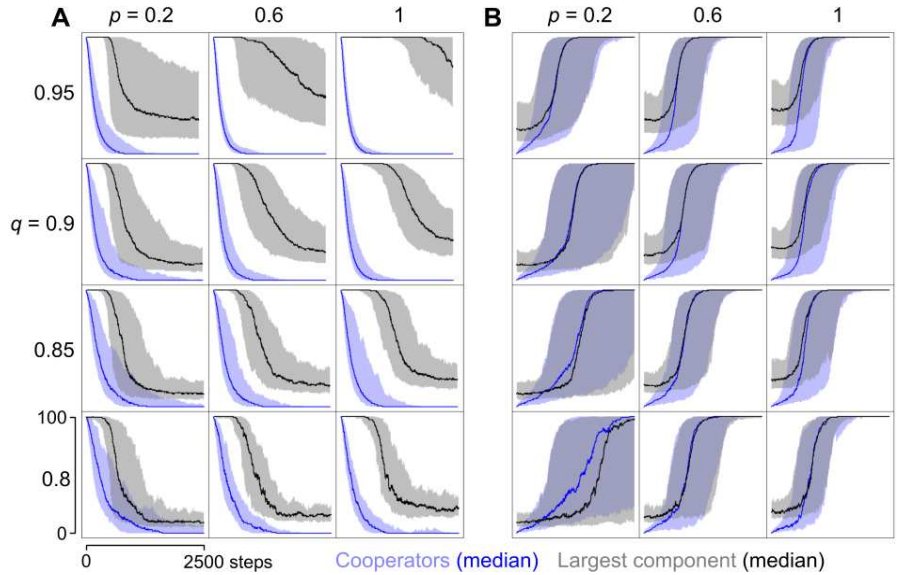

**Fig. 3: Analysis of transitions at various embedding parameters.** Median number of cooperators and size of largest component (dark lines) over time, considering all the transitions observed in individual runs with various combinations of embedding parameters. Other parameters as Fig. 2. The shaded regions represent the 10% (lower bound) and 90% (upper bound) quantiles for the corresponding medians. Consult the Electronic Supplementary Information for the results on the complete range of the embedding parameters.

In Fig. 4 we show how long term cooperation, network structure (long term connectivity and largest component), network stability (number of observed transitions) and long term prosperity depend on the embedding parameters, $p$ and $q$, as well as on the benefit to cost ratio, $b/c$. We observe that the probability $p$ to connect to the role model seems less influential as long as it is above a minimum value close to zero. In contrast, the probability $q$ to connect to the role model's neighbours is crucial for determining the evolution of cooperation, the network structure, stability and prosperity.

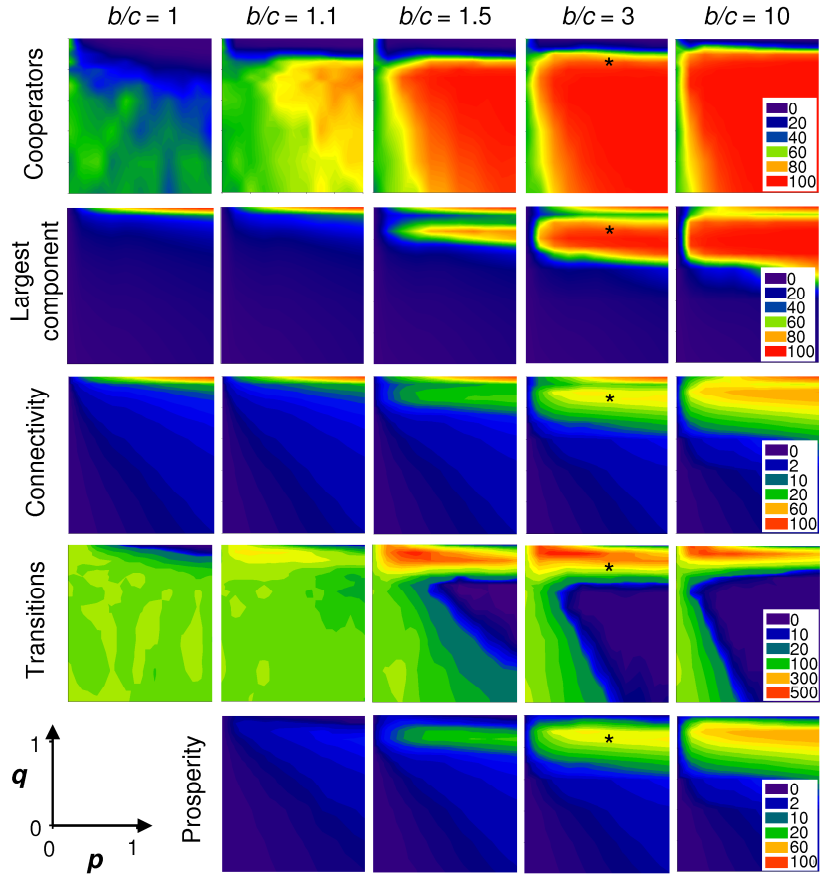

**Fig. 4: Effects of embedding parameters and benefit to cost ratio.** Long term cooperators, largest network component, connectivity, prosperity and number of transitions in relation to embedding parameters, for various benefit to cost ($b/c$) ratios. When $b/c = 1$ long term prosperity is always zero. The black stars in the $b/c = 3$ column denote the $p$, $q$ combination used in Fig. 2. Other parameters as Fig. 2.

The ability for a node to attract newcomers depends on its connectivity but also on its strategy and the strategies of its neighbours. This underlines the co-evolutionary interplay between the spreading of cooperators and network dynamics that leads to a complex trade-off between network stability and long term prosperity. This is illustrated in Fig. 5 for the particular numerical example $b/c = 3$, $p = 0.6$ and varying $q$. With a population of predominantly cooperators, long term connectivity and largest component size increase with increasing $q$ up to peaks where the long term cooperation is close to 100%. Further increasing $q$ allows defectors to invade, leading to a rapid decline in the long term connectivity and size of the largest component. For $q$ close to 1, even defectors form well-connected networks, but with low prosperity. In Fig. 6 we illustrate the topology of networks for a variety of parameters and states of the system. With $q = 0.3$ the network structure (degree and component size distributions) of populations of all-cooperators, all-defectors and mixed states are all similar; there is very low connectivity in all cases. However, for $q = 0.75$ and $q = 0.9$, all-cooperator populations have a much higher connectivity than all-defector populations. There are also interesting differences for mixed populations. For transitions from all-cooperators to all-defectors, we observe that defectors invade a dense network of cooperators. For transitions in the opposite direction, cooperators are seen to form independent clusters with no connections to defectors. For $q = 0.6$ the population of cooperators exists in multiple isolated components, making it difficult for defectors to spread. Here the frequency of transitions is two orders of magnitude lower than for $q = 0.3$ and $q = 0.75$. Thus cooperation is stable, but at the price of low connectivity and low prosperity.

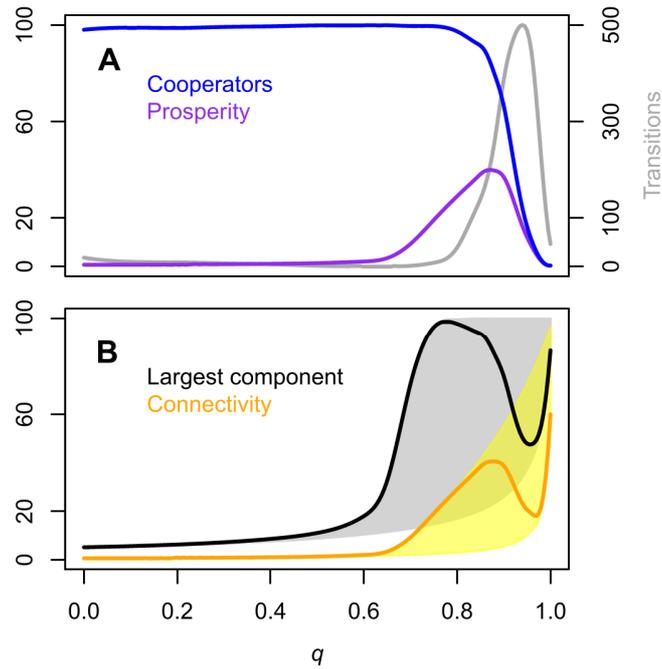

**Fig. 5: Trade-off between network stability and prosperity.** Dependence on $q$ of the long term cooperation, connectivity, largest component, prosperity and number of transitions for $p = 0.6$. Other parameters as Fig. 2. (**A**) Long term cooperators, prosperity and number of transitions seen in $2 \times 10^8$ simulation steps. (**B**) Long term connectivity and largest component plotted against $q$ (solid lines). Shaded areas denote the ranges of connectivity (yellow) and largest component (grey) between all-cooperators (upper boundary) and all-defectors (lower boundary).

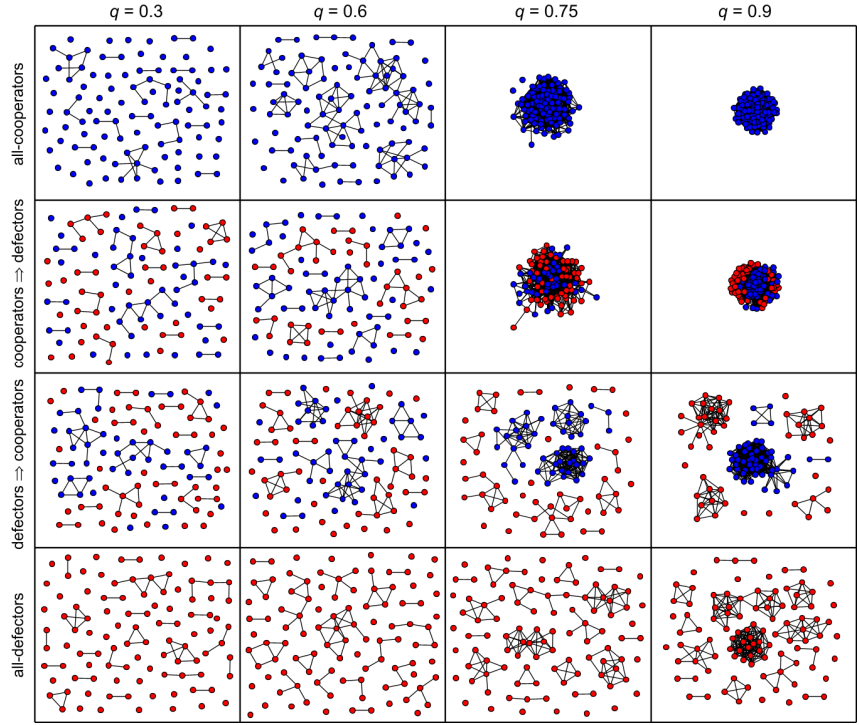

**Fig. 6: Network topology related to *q*.** Typical networks with all-cooperators (top row), all-defectors (bottom row) and the intermediate networks observed during transitions in both directions (middle rows) for $q \in \{0.3, 0.6, 0.75, 0.9\}$. Other parameters as in Fig. 2.

The recurrent formation and fragmentation shown in Fig. 2 can be seen as the result of a conflict between the process of forming clusters and random deletion. Since at each step the node to be removed is chosen uniformly from the population (i.e., not considering the payoff), the expected connectivity of the removed node is equal to the instantaneous average connectivity of the network. As a consequence, the change in the long term connectivity is governed by the rate of the growth process relative to the instantaneous average connectivity of the network. Thus, for network connectivity to increase it is sufficient for newcomers to have higher connectivity than the instantaneous average and not necessary for them to have higher connectivity than the role-model or for the role-model to increase its connectivity. When, by virtue of the strategy mutation rate $u$, a cooperator invades a network of all-defectors, its payoff will be the (equal) lowest in the network and specifically lower than any defectors it is connected to. If by chance the cooperator is chosen as role-model, the newcomer will most likely be a cooperator and, assuming sufficiently large $p$, they will connect and form a pair with higher payoff. Any defectors connected to the cooperators will have higher payoff and this explains why in Fig. 3B we see that invasions by cooperators proceed slowly at first. If the pair of cooperators survive and attract new cooperators, their payoff will eventually be disproportionately greater than the remaining defectors. This then initiates a (probabilistic) positive feedback cycle which causes the synchronous growth of cooperators and connectivity seen in the figures. For $p$ and $q$ not both equal to 1 there is always a non-zero probability that the network will be entirely fragmented (isolated nodes). Thus, for the long term average number of cooperators to be higher than that of defectors, $p$ must be greater than 0 to allow the initial pair of cooperators to form and so have higher payoff than defectors. When, conversely, a defector invades a network of cooperators, it will receive a higher payoff than a cooperator with the same social network and will simultaneously diminish the payoffs of its role-model and its role-model's neighbours. It therefore becomes increasingly likely that a defector will be chosen as a role-model in subsequent steps. The onset of an invasion by defectors is thus rapid, as can be seen in Fig. 3A. In the initial phase of the invasion cooperators are not rare, however the relatively fewer defectors will be disproportionately likely to be chosen as role-models because of their higher payoff. This is illustrated in Fig. S15, where it can be seen that during typical transitions from all-cooperator to all-defector networks with $q = 0.75$ and $q = 0.9$, defectors have comparable or higher average effective payoff than cooperators. During this period the number of defectors increases, but the growth of the connectivity is still affected by the current network connectivity and by the number of cooperators. This explains why there is a delay before the typical collapse in connectivity associated with defectors and why the length of the delay is correlated with $p$ and $q$. As the relative numbers of cooperators thus declines, so too the payoff of the defectors, but now defectors are chosen as role-models by weight of numbers. With zero payoff, the average network connectivity in all-defector networks is at its minimum because the selection of role-models is independent of connectivity.

In the Appendix we provide an analytic theory for the limit of weak selection. We find that the critical benefit-to-cost ratio, beyond which cooperators are favoured over defectors, does not depend on the probability $p$ that newcomers connect to the role model, but is an increasing function of the probability $q$ that the newcomer connects to the role model's neighbours. This agrees with the intuition gained from simulations. Equation 42 in the appendix gives an exact formula that holds for any mutation rate and any population size. For low mutation rate and large population size we find a simple condition for cooperators to prevail, $b/c > (3 + 3\nu + \nu^2) / (2\nu + \nu^2)$, where $\nu = N(1-q)$ is the structural mutation rate (Antal et al., 2009b; Tarnita et al., 2009a), defined as the product of population size and the probability of not adding a link between newcomer and a role model's neighbour. We see that the critical benefit-to-cost ratio approaches one for small values of $q$; here isolated nodes and very small components provide a favourable context for cooperation. For high values of $q$ the critical benefit-to-cost ratio approaches infinity, because the resulting highly connected networks do not allow the evolution of cooperation (Lieberman et al., 2005; Ohtsuki et al., 2006; Szabó and Fáth, 2007). Thus, the weak selection analysis is able to capture the dependence of the critical benefit-to-cost ratio on the parameter $q$ and its independence of $p$, but is not a complete description of the complex evolutionary phenomena observed in the simulations (Nowak et al., 2010a; Traulsen et al., 2010).

## 4 Discussion

Our results show that imitation and varying connectivity constitute a powerful general mechanism for the evolution of cooperation (Nowak, 2006b; Nowak et al., 2010b). We note that this is achieved without the ability of nodes to adjust their strategies or connections (Poncela et al., 2008; Santos et al., 2006), as considered in co-evolutionary networks (Gross and Sayama, 2009; Hanaki et al., 2007; Perc and Szolnoki, 2010). As shown in Fig. 4, already for $b/c = 1.1$ we find a large $p$, $q$-region where the long term cooperation is greater than 75%. For $b/c = 1.5$ there is an even larger $p$, $q$-region which gives a long term cooperation higher than 90%. Cooperators are less abundant than defectors only for very low values of $p$ or for very high values of $q$. If the probability $p$ to connect to the role model is very small, individual cooperators are unlikely to predominate or form connected components.

If the probability $q$ to connect to the role model's neighbours is very large, then the network typically consists of a single highly connected component which behaves like a well-mixed population. In this case defectors dominate.

An intuitive explanation of the described behaviour is that for low $q$ values, cooperators dominate the population, but the network is fragmented; the isolated cooperators do not interact and thus generate low payoff. The prosperity of the network increases as $q$ increases, but if $q$ is too large the network becomes highly connected and cooperators cannot fend off invasion by defectors. Thus, there is an intermediate value of $q$ that maximizes the long term prosperity. Interestingly, as can be observed in Fig. 4 and Fig. 5, the zone of maximum long term prosperity is close to the $q$ value that maximizes the number of transitions between the all-cooperator and all-defector states. In this area of high prosperity the simulations show periods of well-connected networks of cooperators that are frequently interrupted by short-lived transitions to all-defectors (as in Fig. 2). Thus in our model the population is most productive when it is unstable; the long term prosperity is maximized when the frequency of transitions is near its peak. Prosperity increases as more connections between cooperators arise, however as the network becomes more highly connected it begins to resemble a well-mixed population where defectors can take over (Lieberman et al., 2005; Ohtsuki et al., 2006; Szabó and Fáth, 2007). The proliferation of defectors subsequently fragments the network (Fig. 2E-G, 3A), which can then be rapidly rebuilt by a new invasion of cooperators (Fig. 2H, 3B). We note that oscillations between cooperators and defectors have also been observed in other approaches and are a recurrent theme in the evolution of cooperation (Hauert et al., 2006; Nowak and Sigmund, 1989; Wakano et al., 2009).

Our results show that, for dynamic networks, the long term connectivity alone is not an adequate indication of both the level of cooperation and the level of prosperity. This is illustrated in Fig. 4, where it is clear that the average number of cooperators does not follow the trend of connectivity. Moreover, the curve of connectivity shown in Fig. 5B is not monotonic: a single value of connectivity may correspond to three different combinations of cooperation and prosperity. This highlights the fact that the way a network is transformed can strongly affect the spreading of cooperation, obtaining, in a different framework, a result that has been shown for growing networks in (Poncela and et al., 2009). It would be possible to make a quantitative comparison with results obtained on static networks having the same average degree distribution and population ratio as our dynamic networks, however such average networks do not generally correspond to the typical networks seen during simulations, as illustrated in Fig. 2, and such a comparison would be inconclusive.

These results suggest that formation and fragmentation of complex structures (Albert et al., 2000; Barabasi and Albert, 1999; Levin, 2000; Paperin et al., 2011) are correlated and may be a consequence of imitation and internal conflicts between cooperators and defectors; here, the same mechanism that leads to the emergence of a complex network can ultimately cause its fragmentation and allows its subsequent reformation. The presented model is clearly an abstraction of reality, however we note that there are examples of real systems where the collapse and reformation of the network can be plausibly explained by conflicts between cooperators and defectors. For instance, in bacterial communities, which have been considered as networks in (Davies et al., 1998), cooperating cells of *Pseudomonas fluorescens* build biofilms, but mutant cells (defectors) that do not produce the necessary adhesive factors are able to spread, leading to the fragmentation of the structure. The biofilm can then be reformed, under suitable environmental conditions, by the remaining cooperators (Rainey and Rainey, 2003), potentially leading to a cycle of formation and fragmentation. Similar phenomena are observed in the fruiting bodies formed under starvation conditions by cooperative cells of *Myxococcus Xanthus*: defectors invade the population, leading to disruption of the fruiting body structure and possible reconstruction by the cooperative survivors (Travisano and Velicer, 2004). It is also tempting to draw parallels between our results and recent financial crises. These crises (Haldane, 2009a; Haldane and May, 2011; May et al., 2008) have been preceded by a great increase of the financial network connectivity and followed by network fragmentation (Billio et al., 2010; Haldane and May, 2011). The role of imitation and the presence of cooperative and 'greedy' financial institutions have been subjects of the debate on the causes of these crises (Haldane, 2009a).

We have constructed a game theoretic model of dynamic networks able to capture the co-evolutionary interplay between the spreading of cooperators, defectors and the formation and fragmentation of networks. Nodes can be cooperators or defectors and are subject to simple evolutionary criteria: newcomers copy the strategies and connections of successful role-models and old nodes are randomly removed. We have performed simulations and analyses of our model which indicate that it constitutes an effective mechanism for the evolution of cooperation. Moreover, our simulations suggest that endogenous conflicts between cooperators and defectors can cause the periodic formation and fragmentation of complex structures observed in a range of real-world systems. In this light, the prosperity and instability of such complex networks are negatively correlated. While we are aware that there exist many alternatives and potential extensions to our model, we feel that it already captures some of the fundamental mechanisms at work in reality. We believe our findings demonstrate the role and the perils of imitation in the presence of conflicts between cooperators and defectors, suggesting a general trade-off between individual benefit, global welfare and stability in complex networks (Bascompte, 2009; Jackson, 2008; May et al., 2008; Schweitzer et al., 2009; Stiglitz, 2010).

## Acknowledgements

The authors are thankful to Tibor Antal, Péter Csermely, Matthew O. Jackson, Ferenc Jordán and Angel Sánchez for helpful discussions and to Tarcisio Fedrizzi for his initial work on the project. A.C.N. and S.S. gratefully acknowledge support from the Italian Research Fund FIRB (RBPR0523C3) and from Fondazioni CARIPLO and CARITRO, M. C. acknowledges the support of the program JAEDoc15 ("Programa junta para la ampliacion de estudios"), M.A.N. and C.E.T. gratefully acknowledge support from the John Templeton Foundation and the NSF/NIH joint program in mathematical biology (NIH grant R01GM078986).

# Appendix A.

**Analytical solution for the limit of weak selection**

Here we give a complete analytic description of our model for the case of weak selection, $\delta \to 0$.

**A.1 Model description**

We briefly recall here the description of our model. We consider a population of fixed size, $N$, on a dynamic graph. There are two types of individuals, cooperators and defectors. Cooperators pay a cost, $c$, for each neighbour to receive a benefit $b$. Defectors pay no cost and provide no benefits. If, for example, a cooperator is connected to $k$ individuals of whom $j$ are cooperators, then its payoff $= jb - kc$. We use an exponential fitness function. The effective payoff of an individual is $(1+\delta)^{\text{payoff}}$, where $\delta$ is a parameter that scales the intensity of selection.

At any one time step a new individual enters the population and another - randomly chosen - individual exits. This can be done in two ways and we will analyze both. One option is that first someone exits at random and then the newcomer enters; we call this Death-Birth (DB) updating. The other option is that first the newcomer enters and afterwards someone exits; we call this Birth-Death (BD) updating. In the limit of large population size these two processes have the same behavior; however, for small $N$ there are differences between the two processes. For completeness we will do the calculation for both, for exact $N$.

The newcomer is chosen independently from the individual who exits. Thus interactions on our structure are local, but reproduction is global. We will call this global updating.

The newcomer picks one of the existing individuals as a role model. This choice is proportional to the effective payoff. With probability $p$ the newcomer establishes a link to his role model. With probability $q$ the newcomer inherits any one link of the role model. Thus if the role model has $k$ links, then the newcomer inherits all of them with probability $q^k$.

Strategy mutation occurs at rate $u$. With probability $1 - u$ the newcomer adopts the strategy of the role model, but with probability $u$ he adopts the other strategy.

**A.2 Model analysis**

We are studying a Markov process over a state space described as follows. A state $S$ is given by a binary strategy vector $S = (S_1,..., S_n)$ and a binary connection matrix $V = [v_{ij}]$: $s_i$ is the strategy of individual $i$ and it is 1 if $i$ is a cooperator and 0 otherwise; $v_{ij}$ is 1 if $i$ and $j$ are connected and 0 otherwise.

Let $x$ be the frequency of cooperators. We say that on average cooperators are favored over defectors if

$$\langle x \rangle > \frac{1}{2} \qquad (1)$$

where $\langle \cdot \rangle$ denotes the average taken over the stationary distribution of the Markov process. We will now consider how the frequency of cooperators can change from a state to another. There is a change due to selection $\Delta x^{sel}$ and a change due to mutation which on average balance each other. Thus, on average, the total change in the frequency of cooperators is $\langle \Delta x^{tot} \rangle = 0$. Tarnita et al (2009a), Antal et al (2009a; 2009b) have shown that for global updating, the condition (1) that cooperators are favored over defectors is equivalent to asking that the average change due to selection in the frequency of cooperators is positive. In other words, cooperation wins if on average selection favors it:

$$\langle \Delta x^{sel} \rangle > 0. \tag{2}$$

We can explicitly write the average over the stationary distribution as

$$\langle \Delta x^{sel} \rangle = \sum_S \Delta x_S^{sel} \pi_S \tag{3}$$

Here $\Delta x_S^{sel}$ is the change due to selection in state $S$ and $\pi_S$ is the probability that the system is in state $S$. Since we are interested in the results obtained in the weak selection limit, $\delta \to 0$, we only need to work with the constant and first-order terms in $\delta$ of the expression (2). The constant term is the average change in the frequency of cooperators at neutrality, which is zero. Using our assumption that the transition probabilities are analytic at $\delta = 0$ we can conclude as in Tarnita et al (2009b) that the probabilities $\pi_S$ and the change due to selection in each state $\Delta x_s$ are also analytic at $\delta = 0$. Hence we can write the first order Taylor expansion of the average change due to selection at $\delta = 0$:

$$\langle \Delta x^{sel} \rangle \approx \delta \left( \sum_S \left. \frac{\partial \Delta x_S^{sel}}{\partial \delta} \right|_{\delta=0} \pi_S(\delta = 0) + \sum_S \Delta x_S^{sel}(\delta = 0) \left. \frac{\partial \pi_S}{\partial \delta} \right|_{\delta=0} \right) \tag{4}$$

In particular, since we are only dealing with global updating with constant death (individuals are replaced at random with probability $1/N$), the change in frequency at neutrality in each state is zero. Thus the second term in (4) is zero and hence, in the limit of weak selection, condition (2) becomes

$$\langle \Delta x^{sel} \rangle \approx \sum_S \left. \frac{\partial \Delta x^{sel}_S}{\partial \delta} \right|_{\delta=0} \pi_S(\delta=0) := \left\langle \left. \frac{\partial \Delta x^{sel}}{\partial \delta} \right|_{\delta=0} \right\rangle_0 > 0 \qquad (5)$$

Here := denotes notation; $\langle \cdot \rangle_0$ denotes the average over the stationary distribution at neutrality, $\delta = 0$. It is weighted by the probability $\pi_S$ ($\delta = 0$) that the system is in state $S$ at neutrality. In other words, in the limit of weak selection, the condition that the average change due to selection is greater than zero is equivalent to the condition that the neutral average of the first derivative with respect to $\delta$ of the change due to selection is greater than zero.

Next we can explicitly write the expected change due to selection in a certain state as

$$\Delta x^{sel} = \sum_i s_i(w_i - 1) \qquad (6)$$

where $w_i$ is the expected number of offspring of individual $i$. We are dealing with a Moran process with global updating and hence we can write

$$w_i = 1 - \frac{1}{N} + \frac{f_i}{\sum_j f_j} \qquad (7)$$

This is because each individual survives with probability $1 - 1/N$ and gives birth with probability proportional to his payoff. In our model, the effective payoff is given by the exponential function $(1+\delta)^{\text{payoff}}$; however, in the limit of weak selection, this becomes $1+\delta\text{payoff}$ and hence we can write the effective payoff of individual $i$ as

$$f_i = 1 + \delta \sum_j v_{ij}(-cs_i + bs_j) \tag{8}$$

Here and throughout we assume that there are no self-interactions.

Substituting (7) and (8) into (6) and taking the limit of weak selection we obtain

$$\begin{aligned}\Delta x^{sel} &= \delta\Big[\sum_i s_i\Big[\sum_j v_{ij}(-cs_i + bs_j) - \frac{1}{N}\sum_j\sum_k v_{jk}(-cs_j + bs_k)\Big]\Big] \\ &= \delta\Big[b\Big(\sum_{i,j} s_i s_j v_{ij} - \frac{1}{N}\sum_{i,j,k} s_i s_k v_{ij}\Big) - c\Big(\sum_{i,j} s_i v_{ij} - \frac{1}{N}\sum_{i,j,k} s_i s_k v_{ij}\Big)\Big]\end{aligned} \tag{9}$$

Using (9) into (5) we obtain the condition for cooperators to be favored over defectors to be

$$\frac{b}{c} > \frac{\langle \sum_{i,j} s_i v_{ij}\rangle_0 - \frac{1}{N}\langle \sum_{i,j,k} s_i s_k v_{ij}\rangle_0}{\langle \sum_{i,j} s_i s_j v_{ij}\rangle_0 - \frac{1}{N}\langle \sum_{i,j,k} s_i s_k v_{ij}\rangle_0} \tag{10}$$

The critical benefit to cost ratio in (10) can be rewritten as follows

$$\frac{b}{c} > \frac{1 - \overline{G}}{G - \overline{G}} \tag{11}$$

where

$$\begin{aligned} G &= Pr_0(s_i = s_j | v_{ij} = 1) \\ \overline{G} &= Pr_0(s_j = s_k | v_{ij} = 1) \end{aligned} \tag{12}$$

The notation $Pr_0$ means that the probabilities are calculated at neutrality. However, for simplicity we will use the notation $Pr$ from now on. To define $G$ and $\bar{G}$ we pick three individuals *i, j, k* at random with replacement such that *i* and *j* are connected. Given this choice, $G$ is the probability that $i$ and $j$ have the same strategy and $\bar{G}$ is the probability that $j$ and $k$ have the same strategy. In other words, $G$ is the probability that two individuals that are connected also have the same strategy, whereas $\bar{G}$ is the probability that two random individuals have the same strategy, modified to account for the fact that the structure is dynamical. We will proceed to calculate these quantities below.

### A.3 Calculating $G$ and $\bar{G}$

For simplicity, we want to calculate quantities where the three individuals are chosen without replacement. Let us make the following notation

$$z = \Pr(v_{ij} = 1 \mid i \neq j) \tag{13}$$

$$g = \Pr(v_{ij} = 1 \text{ and } s_i = s_j \mid i \neq j) \tag{14}$$

$$h = \Pr(v_{ij} = 1 \text{ and } s_i = s_k \mid i \neq j \neq k) \tag{15}$$

Then the critical benefit-to-cost ratio (11) can be expressed in terms of $z$, $g$ and $h$ as

$$\left(\frac{b}{c}\right)^* = \frac{(N-2)(z-h) + z - g}{(N-2)(g-h) - z + g} \tag{16}$$

In the large $N$ limit we have

$$\left(\frac{b}{c}\right)^* = \frac{z-h}{g-h} \tag{17}$$

Here for simplicity we use the same notation, but by *z*, *g* and *h* we mean their large *N* limits.

Thus, for calculating the critical benefit-to-cost ratio in the limit of weak selection, it suffices to find *z*, *g* and *h* in the neutral case: *z* is the probability that two distinct randomly picked individuals are connected; *g* is the probability that they are connected and have the same strategy. For *h* we need to pick three distinct individuals at random; then *h* is the probability that the first two are connected and the latter two have the same strategy.

In general these quantities cannot be written as independent products of the probability of being connected times the probability of having the same strategy. However, if we fix the time to their most recent common ancestor (MRCA) and we take the limit of large *N*, these quantities become independent (Wakeley, 2008).

Two individuals always have a common ancestor if we go back in time far enough. However, we cannot know how far we need to go back. Thus, we have to account for the possibility that *t* takes values anywhere between 1 and ∞. Note that $t = 0$ is excluded because we assume that the two individuals are distinct. Moreover, we know that this time is affected neither by the strategies, nor by the set memberships of the two individuals. It is solely a consequence of the Moran dynamics.

### A.3.1 Probability of given coalescence time

In what follows, for simplicity of the exposition we will do the calculation for BD updating and, where different, we will specify in footnotes what the corresponding quantities are for DB updating.

We first find the probability that the two individuals coalesced in time $t = 1$. This probability differs between the two processes. Thus, for BD updating[4] we must have that one of them is the parent and the other is the offspring; moreover, we have to make sure that the parent has not died in the last update step. Hence the probability that they coalesced in time $t = 1$ is $2/N^2$ which gives

$$P(t) = \left(1 - \frac{2}{N^2}\right)^{t-1} \frac{2}{N^2} \qquad (18)$$

Similarly, we can write the probability that three individuals coalesce such that the first two have a MRCA $t_3$ update steps backward and this MRCA and the third individual require $t_2$ more update steps to coalesce.

For BD updating[5], this probability is given by

$$\Pr(t_3, t_2) = \left(1 - \frac{6}{N^2}\right)^{t_3 - 1} \frac{6}{N^2} \left(1 - \frac{2}{N^2}\right)^{t-1} \frac{2}{N^2} \qquad (19)$$

---

[4] For DB updating we must have that one of them is the parent and the other is the offspring, which happens with probability $2/[N(N-1)]$. Then we can write that

$$P(t) = \left(1 - \frac{2}{N(N-1)}\right)^{t-1} \frac{2}{N(N-1)}$$

[5] For DB updating we have

$$\Pr(t_3, t_2) = \left(1 - \frac{6}{N(N-1)}\right)^{t_3 - 1} \frac{6}{N(N-1)} \left(1 - \frac{2}{N(N-1)}\right)^{t_2 - 1} \frac{2}{N(N-1)}$$

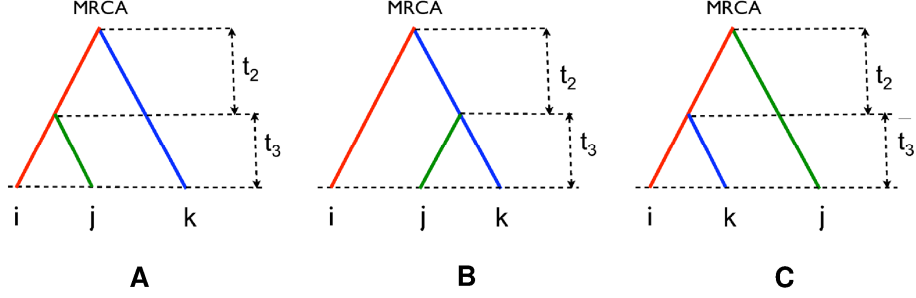

**Fig. A1: There are three possibilities for the ancestry of three individuals**: (a) *i* and *j* coalesce first and then they coalesce with *k*; (b) *j* and *k* coalesce first and then they coalesce with *i*; (c) *i* and *k* coalesce first and then they coalesce with *j*. Each case happens with probability 1/3.

If we introduce a rescaled time $\tau_* = t_*/(N^2/2)$ and consider the continuous-time process in the limit of *N* large we obtain the probability density functions which are identical for both DB and BD

$$p(\tau) = e^{-\tau}$$
$$p(\tau_3, \tau_2) = 3e^{-3\tau_3 - \tau_2} \quad (20)$$

### A.3.2 Probability that two individuals have the same strategy at time *T = t* from the MRCA

Let $P_S(t)$ be the probability that two individuals have the same strategy at time *T = t* from the MRCA. At time *T* = 1 we have $P_S(1) = 1 - u$. In general, the probability that two individuals have the same strategy at time *T = t* is the probability that their ancestors had the same strategy in the previous step, at time *T = t* − 1 plus/minus what is gained/lost by mutation if there was a reproductive step in their ancestry lines from time *t* - 1 to time *t*. That is

$$P_s(t) = P_s(t-1)(1 - P_{B2} + P_{B2}(1-u)) + (1 - P_s(t-1))uP_{B2} \quad (21)$$

where $P_{B2}$ is the probability that a birth event happened in the ancestry lines of two individuals in the previous update step. It easily follows that the probability that two individuals have the same strategy at time $T = t$ from the MRCA is

$$P_s(t) = \frac{1}{2} + \frac{1-2u}{2}\left(1 - 2uP_{B2}\right)^{t-1} \tag{22}$$

For BD updating[6] it is easy to see that $P_{B2} = 2(N-1)/(N^2-2)$.

For the continuous time process, letting $\tau = t/(N^2/2)$ we obtain the density function

$$p_s(\tau) = \frac{1 + e^{-\mu\tau}}{2} \tag{23}$$

where $\mu = 2Nu$. Note that we are taking the limits of large $N$ and small $u$ at the same time, such that $\mu = 2Nu$ is a well-defined quantity.

### A.3.3 Probability that two individuals are connected at time $T = t$ from the MRCA

Let $P_C(t)$ be the probability that two individuals are connected at time $T = t$ from the MRCA. Clearly at time $T = 1$ we have $P_C(1) = p$. In general, the probability that two individuals are connected at time $T = t$ after their MRCA is the same as the probability that their ancestors were connected at time $T = t - 1$ multiplied by the probability that in the subsequent update step they stayed connected (either because neither of them was picked

---

[6] For DB updating, the probability $P_{B2}$ is the probability of picking in the previous update step a death-birth pair such that neither of the two dies but one of them gives birth. Thus $P_{B2} = 2(N-2)/[N(N-1)-2] = 2/(N+1)$ for DB updating. The recurrence relation is identical.

for reproduction or, if either was picked the offspring established a connection). Thus, we have

$$P_c(t) = P_c(t-1)((1 - P_{B2}) + qP_{B2}) \tag{24}$$

where $P_{B2}$ is as before, the probability that a birth event happened in the ancestry lines of two individuals in the previous update step. Thus we find that

$$P_c(t) = p\left(1 - (1-q)P_{B2}\right)^{t-1} \tag{25}$$

For the continuous time process, letting $\tau = t/(N^2/2)$ we obtain the density function

$$p_c(\tau) = pe^{-\nu\tau} \tag{26}$$

where $\nu = N(1-q)$. As before, this quantity is meaningful as it is taken for the limit of large $N$ and large $q$.

Note that if at time $T = 1$ after the MRCA two individuals are not connected, then their offspring will not be connected no matter what. However, after $T = 1$ all that matters is the probability $q$ that offspring add links to their parents' neighbours.

### A.3.4 Critical benefit-to-cost ratio for $N$ large

As discussed in Wakekey (2008), Antal et al. (2009b) and Tarnita et al. (2009a), in the limit of large population size the probability that two individuals are connected and have the same strategy at time $\tau$ after the MRCA is a product of the respective independent probabilities. In this case we can write

$$z = \int_0^\infty p_c(\tau)p(\tau)\, d\tau = \frac{p}{1+\nu}$$

$$g = \int_0^\infty p_c(\tau)p_s(\tau)p(\tau)\, d\tau = \frac{p(2+2\nu+\mu)}{2(1+\nu)(1+\nu+\mu)}$$

$$h = \frac{1}{3}\int_0^\infty d\tau_3 \int_0^\infty d\tau_2\, [p_s(\tau_3)p_c(\tau_3+\tau_2) + p_s(\tau_3+\tau_2)p_c(\tau_3) + p_s(\tau_3+\tau_2)p_c(\tau_3+\tau_2)]p(\tau_3,\tau_2)$$

$$= \frac{p(\mu^3 + 2\mu^2(3+\nu) + 2(1+\nu)(3+\nu) + \mu(11+\nu(9+\nu)))}{2(1+\mu)(1+\nu)(1+\mu+\nu)(3+\mu+\nu)}$$

(27)

where $\mu = 2Nu$ and $\nu = N(1-q)$.

Using (17) we can calculate the critical benefit to cost ratio to be

$$\left(\frac{b}{c}\right)^* = \frac{z-h}{g-h} = \frac{3+\mu^2+2\mu(2+\nu)+\nu(3+\nu)}{\nu(2+\mu+\nu)} \tag{28}$$

This result holds for both DB and BD updating. In the limit of low strategy mutation, the benefit-to-cost ratio simplifies to

$$\left(\frac{b}{c}\right)^* = \frac{3+3\nu+\nu^2}{\nu(2+\nu)} \tag{29}$$

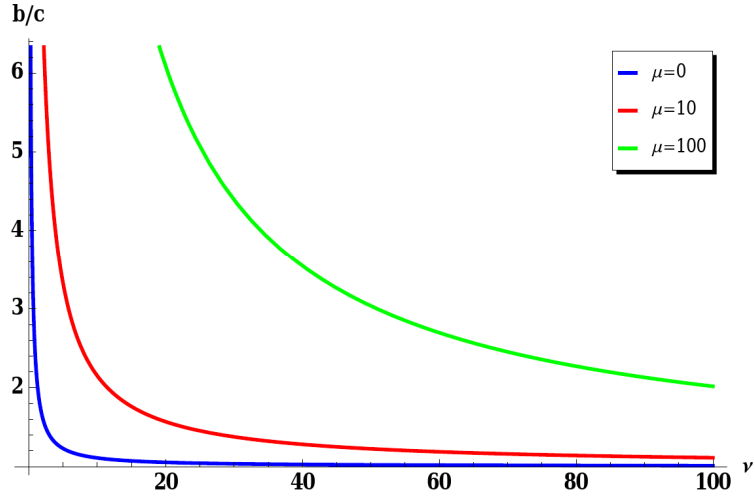

**Fig. A2: Critical benefit-to-cost ratio as a function of the effective connection mutation rate $v = N(1-q)$.** The effective strategy mutation rate is $\mu = 0$, 10 and 100. The origin of the axes is (0,1).

Finally, using the result in Tarnita et al. (2009b) we can calculate the structure coefficient σ

$$\sigma = \frac{(b/c)^* + 1}{(b/c)^* - 1} = 2\nu - 1 + \frac{6}{3+\nu} \tag{30}$$

### A.4 Critical benefit-to-cost ratio for exact $N$

For exact $N$, the probabilities above are not independent. Hence, we need to calculate directly the probability that two individuals are connected and have the same strategy at time $t$ after the MRCA. Similarly for the other quantities.

#### A.4.1 Probability that two individuals are connected

First we calculate the probability $z$ that two individuals are connected. This follows directly from our derivation above, using (24)

$$z = \sum_{t=1}^{\infty} P_c(t) P(t) \tag{31}$$

For BD updating[7] we find

$$z = \frac{p}{N(1-q)+q} \qquad (32)$$

**A.4.2 Probability that two individuals are connected and have the same strategy**

Let $P_{cs}(t)$ be the probability that two individuals are connected and have the same strategy at time $t$ after the MRCA. Then $P_{cs}(1) = p(1-u)$. In general, for two individuals to be connected and have the same strategy at time $t$ it is necessary that their ancestors at time $t-1$ were connected but it is not necessary that they had the same strategy. Letting $P_{c\bar{s}}(t)$ be the probability that two individuals are connected but do not have the same strategy at time $t$ we can write

$$\begin{aligned}P_{cs}(t) &= P_{cs}(t-1)(1-P_{B2}+P_{B2}(1-u)q) + P_{c\bar{s}}(t-1)P_{B2}qu \\ P_{c\bar{s}}(t) &= P_{c\bar{s}}(t-1)(1-P_{B2}+P_{B2}(1-u)q) + P_{cs}(t-1)P_{B2}qu\end{aligned} \qquad (33)$$

Solving the recurrences with base cases $P_{cs}(1) = p(1-u)$ and $P_{c\bar{s}}(1) = pu$ we obtain

$$P_{cs}(t) = \frac{p}{2}\Big(\big(1-(1-q)P_{B2}\big)^{t-1} + (1-2u)\big(1-(1-q+2qu)P_{B2}\big)^{t-1}\Big) \qquad (34)$$

Here the recurrence is the same for both DB and BD updating; the only difference is in the value of $P_{B2}$ as specified before. To find $g$ we then need to calculate the infinite sum

---

[7] For DB updating we find

$$z = \frac{p}{N(1-q)-1+2q}$$

$$g = \sum_{t=1}^{\infty} P_{cs}(t) P(t) \tag{35}$$

For BD updating[8] we find :

$$g = \frac{p(2q(-1+2u) + N(-2-2q(-1+2u)+2u))}{2(N(-1+q)-q)(N+q-Nq+2(-1+N)qu)} \tag{36}$$

**A.4.3 Probability that first two are connected and latter two have same strategy**

This calculation is along the same lines as above. However, now we need to take into account the three coalescent probabilities (as in Figure S4). Each one of them happens with probability 1/3. Let $P_{CS}(t)$ be the probability that given three random individuals, the first two are connected and the latter two have the same strategy. Let $P_{C\bar{S}}(t)$ be the probability that the first two are connected but the latter two do not have the same strategy. Then one can write

$$\begin{aligned} P_{CS}(t_2, t_3) &= P_{CS}(t_2, t_3-1)(1-P_{B3} + P_{B3}\frac{1}{3}(q+1-u+q(1-u))) + \\ &\quad + P_{C\bar{S}}(t_2, t_3-1) P_{B3}\frac{1}{3}u(1+q) \\ P_{C\bar{S}}(t_2, t_3) &= P_{C\bar{S}}(t_2, t_3-1)(1-P_{B3} + P_{B3}\frac{1}{3}(q+1-u+q(1-u))) + \\ &\quad + P_{CS}(t_2, t_3-1) P_{B3}\frac{1}{3}u(1+q) \end{aligned} \tag{37}$$

---

[8] For DB updating we find

$$g = \frac{p(2 - 4q(1-2u) - 2u + N(-2-2q(-1+2u)+2u))}{2(1+N(-1+q)-2q)(-1+N+2q-Nq+2(-2+N)qu)}$$

Here $P_{B3}$ is the probability that there was a birth event in the ancestry lines of the three individuals.

For BD updating[9] we have $P_{B3} = 3(N-2)/(N^2-6)$.

Next we need to write the base case recurrences. These depend on which of the three cases in Figure S4 we are in. Thus we have

- if we are in case (a), such that individuals *i* and *j* coalesced first and then they coalesced with *k* then

$$\begin{aligned}P_{CS}(t_2+1,0) &= P_s(t_2)\frac{p}{2}(2-u) + (1-P_s(t_2))\frac{p}{2}u \\ P_{C\bar{S}}(t_2+1,0) &= (1-P_s(t_2))\frac{p}{2}(2-u) + P_s(t_2)\frac{p}{2}u\end{aligned} \quad (38)$$

where $P_s(t)$ is, as before, the probability that two individuals have the same strategy at time *t* after their MRCA.

- if we are in case (b), such that individuals *j* and *k* coalesced first and then they coalesced with *i* then

$$\begin{aligned}P_{CS}(t_2+1,0) &= P_c(t_2)\frac{1}{2}(1-u)(1+q) \\ P_{C\bar{S}}(t_2+1,0) &= P_c(t_2)\frac{1}{2}u(1+q)\end{aligned} \quad (39)$$

where $P_c(t)$ is, as before, the probability that two individuals are connected at time *t* after their MRCA.

---

[9] For DB updating we have $P_{B3} = 3/(N+2)$

- if we are in case (c), such that individuals *i* and *k* coalesced first and then they coalesced with *j* then

$$P_{CS}(t_2+1,0) = P_{cs}(t_2)\frac{1}{2}(q+1-u) + P_{c\bar{s}}(t_2)\frac{u}{2}$$
$$P_{C\bar{S}}(t_2+1,0) = P_{c\bar{s}}(t_2)\frac{1}{2}(q+1-u) + P_{cs}(t_2)\frac{u}{2}$$
(40)

where $P_{cs}(t)$ and $P_{c\bar{s}}(t)$ are, as before, the probability that two individuals are connected and have the same strategy at time *t* after their MRCA, respectively that they are connected but do not have the same strategy.

Performing this calculation for BD updating[10] we obtain

$h = numerator/denominator$:

---

[10] For DB updating we obtain

$$\begin{aligned}
numerator =\ & p(-2(N(-1+q)-3q)(1+N(-1+q)-2q) + (-7-2N^3(-1+q)^2+ \\
& + 6N^2(-1+q)(-1+4q) + q(-15+76q) + N(3+(53-78q)q))u+ \\
& + 2(12 - N(19+(-8+N)N) - 18q + N(-1+(9-2N)N)q+ \\
& + (-2+N)(36+N(-23+3N))q^2)u^2 \\
& - 4(-2+N)(1+q)(-1+N+(-5+N)(-2+N)q)u^3) \\
denominator =\ & (2(1+N(-1+q)-2q)(1+2(-2+N)u)(-1+N+2q-Nq+ \\
& + 2(-2+N)qu)(N+3q-Nq+(-3+N)(1+q)u))
\end{aligned}$$

$$\begin{aligned}
numerator =\ & p(2q(1-2u)^2(-1+2q(-1+u)+2u) - N(-1+2u)(2(-1+q)(1+3q)+ \\
& + (5-3q(1+8q))u + 2(1+q)(-1+9q)u^2) + 2N^2(-(-1+q)^2 + 9(-1+q)qu+ \\
& + (5+(3-20q)q)u^2 + 2(1+q)(-1+6q)u^3) - \\
& - 2N^3 u(1+q(-1+u)+u)(1+q(-1+2u))) \\
denominator =\ & 2(N(-1+q)-q)(1+2(-1+N)u)(N+q-Nq+ \\
& + 2(-1+N)qu)(1+N+2q-Nq+(-2+N)(1+q)u)
\end{aligned}$$
(41)

### A.4.4 Benefit-to-cost ratio for exact N

Using formula (16) together with (32), (36) and (41) we obtain the exact critical benefit-to-cost ratio for BD updating[11] to be $(b/c)^* = num/den$ where

$$\begin{aligned}
num =\ & -2q(-1+2u)(-1+2q(-1+u)+2u) + 2N^3(1+q(-1+u)+u)(1+q(-1+2u)) - \\
& - 2N^2(1+u+q(-5+4q+3(1-4q)u+8(1+q)u^2)) + \\
& + N(-1+2u+q(-3-8u+10(q-3qu+2(1+q)u^2))) \\
den =\ & 2N^3(-1+q)(1+q(-1+u)+u)(-1+2u) - 2q(-1+2u)(-1+2q(-1+u)+2u) - \\
& - 4N^2(1+q(-3+2q)+u+(5-6q)qu+(-3+q+4q^2)u^2) + \\
& + N(-3+2(5-4u)u+10q^2(-1+u)(-1+2u)+q(-7+4u(1+3u)))
\end{aligned}$$

(42)

---

[11] For DB updating we have

$$\begin{aligned}
num =\ & -2+7N-6N^2+2N^3+12q-25Nq+18N^2q-4N^3q-16q^2+24Nq^2- \\
& - 12N^2q^2+2N^3q^2+2(N(3+(-3+N)N)+N(7+N(-7+2N))q- \\
& - 3(-2+N)^3q^2-3(1+q))u+4(-2+N)(1+q)(-1+(-2+N)^2q)u^2 \\
den =\ & -2+7N-8N^2+2N^3+12q-31Nq+20N^2q-4N^3q-16q^2+24Nq^2- \\
& - 12N^2q^2+2N^3q^2-2(3-6N+N^3+(3-2N(11+N(-9+2N)))q+ \\
& + 3(-2+N)^3q^2)u+4(-2+N)(1+q)(-1+3N-N^2+(-2+N)^2q)u^2
\end{aligned}$$

**A.5 Comparison with neutral simulations**

In this section we use the numerical simulation method developed by Nathanson et al. (2009) to verify the accuracy of our calculations. Tarnita et al. (2009b) show that for any structured population, under very mild assumptions, the weak selection condition for strategy cooperators to be favoured over defectors is given by

$$\frac{b}{c} > \frac{\sigma + 1}{\sigma - 1} \qquad (43)$$

For global updating with constant death or birth rates, Nathanson et al (2009) find an expression for the structure coefficient σ which enables us to perform very fast and accurate numerical simulations. For each state of the system, let $N_A$ be the number of individuals using strategy *A*; the number of individuals using strategy *B* is $N_B = N - N_A$. Furthermore, let $I_{AA}$ denote the total number of encounters that *A* individuals have with other *A* individuals. Note that every *AA* pair is counted twice because each *A* individual in the pair has an encounter with another *A* individual. Let $I_{AB}$ denote the total number of interactions that an *A* individual has with *B* individuals. Then the structure coefficient, $\sigma$, can be written as

$$\sigma = \frac{\langle I_{AA} N_B \rangle_0}{\langle I_{AB} N_B \rangle_0} \qquad (44)$$

This suggests a simple numerical algorithm for calculating this quantity for our spatial process. We simulate the process under neutral drift for many generations. For each state we evaluate $N_B$, $I_{AA}$, and $I_{AB}$. We add up all $N_B I_{AA}$ products to get the numerator in equation (3), and then we add up all $N_B I_{AB}$ products to get the denominator. We obtain the perfect agreement in Figure A3.

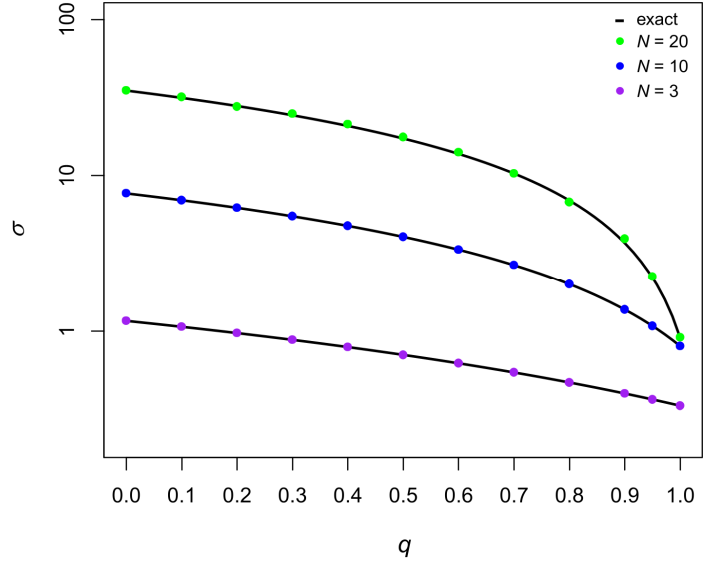

**Fig. A3: Comparison of $\sigma$ with numerical simulations for various *N, q* and *u*.** The analytical curves show close agreement with the simulated data points. Each point was generated by averaging statistics taken from two simulation runs of $10^9$ steps, ignoring the first $10^7$ steps, with $p = 0.5$. The mutation rates used were $u = 0.1$ for $N = 3$, $u = 0.05$ for $N = 10$, $u = 0.01$ for $N = 20$.

**A.6 Prosperity**

In this section we calculate the average wealth of the population for weak selection. Let $F$ be the total effective payoff of the population after taking the limit of weak selection. It can be written as $F = N + \delta P$ where $P$ is the total payoff in the population. What we want to maximize is $W = \langle F \rangle = N + \delta \langle P \rangle$ which is the average prosperity. The total payoff in the population in a state $S$ can be written as

$$P = \sum_i \sum_j v_{ij}(-cs_i + bs_j) \qquad (45)$$

Thus the prosperity becomes

$$\langle W \rangle = N + \delta \langle \sum_i \sum_j v_{ij}(-cs_i + bs_j) \rangle$$
$$= N + \delta(b-c) \langle \sum_i \sum_j s_i v_{ij} \rangle \quad (46)$$
$$= N + \delta(b-c) N^2 \Pr(s_i = 1 \text{ and } v_{ij} = 1)$$

Thus what needs to be maximized is the average probability at neutrality that two random individuals are connected and the first one is a cooperator. This turns out to be

$$Pr(v_{ij} = 1) = \frac{p}{1+\nu} \quad (47)$$

Thus, for weak selection, the prosperity of the system increases with $q$, which is a result we observe in the simulations. However, what we do not find in our calculation for weak selection is an optimum intermediate $q$ which maximizes the prosperity. This is because at neutrality this calculation does not capture the clustering behavior of cooperators as opposed to the dispersing behavior of defectors because at neutrality they are interchangeable labels. As the intensity of selection is increased the probability of being connected reflects more and more the clustering effect. Below we give the plot of this probability for weak selection.

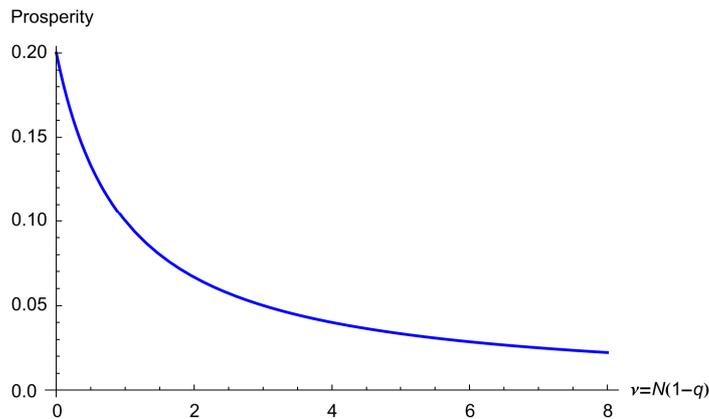

**Fig. A4**: **Prosperity as a function of $\nu = N(1-q)$ for large $N$ and $p = 0.2$.**

# References

Akerlof, G.A., and Shiller, R.J., 2009. Animal spirits : how human psychology drives the economy, and why it matters for global capitalism. Princeton University Press, Princeton.

Albert, R., Jeong, H., and Barabasi, A.-L., 2000. Error and attack tolerance of complex networks. Nature 406, 378-382.

Antal, T., Traulsen, A., Ohtsuki, H., Tarnita, C.E., and Nowak, M.A., 2009a. Mutation-selection equilibrium in games with multiple strategies. J Theor Biol 258, 614-22.

Antal, T., Ohtsuki, H., Wakeley, J., Taylor, P.D., and Nowak, M.A., 2009b. Evolution of cooperation by phenotypic similarity. Proc Natl Acad Sci U S A 106, 8597-600.

Aviles, L., 1999. Cooperation and non-linear dynamics: An ecological perspective on the evolution of sociality. Evol Ecol Res 1, 459-477.

Bandura, A., 1985. Social Foundations of Thought and Action: A Social Cognitive Theory. Prentice Hall.

Barabasi, A.-L., and Albert, R., 1999. Emergence of Scaling in Random Networks. Science 286, 509-512.

Bascompte, J., 2009. Disentangling the web of life. Science 325, 416-9.

Billio, M., Getmansky, M., Lo, A.W., and Pelizzon, L., 2010. Econometric Measures of Systemic Risk in the Finance and Insurance Sectors. NBER Working Paper.

Boccaletti, S., Latora, V., Moreno, Y., Chavez, M., and Hwang, D.U., 2006. Complex networks: Structure and dynamics. Phys Rep 424, 175-308.

Bonabeau, E., 2004. The perils of the imitation age. Harv Bus Rev 82, 45-54.

Castellano, C., Fortunato, S., and Loreto, V., 2009. Statistical physics of social dynamics. Rev Mod Phys 81, 591.

Csermely, P., 2009. Weak Links: The Universal Key to the Stability of Networks and Complex Systems Springer, Heidelberg.

Davidsen, J., ouml, rn, Ebel, H., and Bornholdt, S., 2002. Emergence of a Small World from Local Interactions: Modeling Acquaintance Networks. Physical Review Letters 88, 128701.

Davies, D.G., Parsek, M.R., Pearson, J.P., Iglewski, B.H., Costerton, J.W., and Greenberg, E.P., 1998. The Involvement of Cell-to-Cell Signals in the Development of a Bacterial Biofilm. Science 280, 295-298.

Dorogovtsev, S.N., and Mendes, J.F.F., 2003. Evolution of Networks: From Biological Nets to the Internet and WWW (Physics). Oxford University Press, Inc.

Erdős, P., and Rényi, A., 1960. On the evolution of random graphs. Publ. Math. Inst. Hungar. Acad. Sci 5, 17-61.

Gross, T., and Sayama, H., 2009. Adaptive Networks. Springer, Heidelberg.

Haldane, A.G., 2009a. Credit is trust. Speech given at the Association of Corporate Treasurers, Leeds.

Haldane, A.G., 2009b. Rethinking the financial network. Speech delivered at the Financial Student Association, Amsterdam.

Haldane, A.G., and May, R.M., 2011. Systemic risk in banking ecosystems. Nature 469, 351-355.

Hanaki, N., Peterhansl, A., Dodds, P.S., and Watts, D.J., 2007. Cooperation in Evolving Social Networks. Management Sci 53, 1036-1050.

Hauert, C., Holmes, M., and Doebeli, M., 2006. Evolutionary games and population dynamics: maintenance of cooperation in public goods games. Proc Biol Sci 273, 2565-70.

Helbing, D., 2010. Systemic risks in society and economics. International Risk Governance Council.

Hofbauer, J., and Sigmund, K., 1988. Evolutionary games and population dynamics. Cambridge University Press, Cambridge.

Jackson, M.O., 2008. Social and Economic Networks. Princeton University Press, Princeton.

Jackson, M.O., and Rogers, B.W., 2007. Meeting Strangers and Friends of Friends: How Random Are Social Networks? Amer Econ Rev 97, 890-915.

Kleinberg, J.M., Kumar, R., Raghavan, P., Rajagopalan, S., and Tomkins, A.S., The web as a graph: measurements, models, and methods, Proceedings of the 5th annual international conference on Computing and combinatorics, Springer-Verlag, Tokyo, Japan 1999.

Krapivsky, P.L., and Redner, S., 2005. Network growth by copying. Phys Rev E Stat Nonlin Soft Matter Phys 71, 036118.

Kumar, R., Raghavan, P., Rajagopalan, S., Sivakumar, D., Tomkins, A., and Upfal, E., Stochastic models for the web graph, 41st Annual Symposium on Foundations of Computer Science, Proceedings, 2000, pp. 57-65.

Levin, S.A., 2000. Fragile Dominion. Helix Books, Santa Fe, NM, USA.

Lieberman, E., Hauert, C., and Nowak, M.A., 2005. Evolutionary dynamics on graphs. Nature 433, 312-316.

May, R.M., Levin, S.A., and Sugihara, G., 2008. Complex systems: ecology for bankers. Nature 451, 893-5.

Montoya, J.M., Pimm, S.L., and Sole, R.V., 2006. Ecological networks and their fragility. Nature 442, 259-64.

Moran, P.A.P., 1962. The statistical processes of evolutionary theory. Clarendon Press, Oxford.

Nathanson, C.G., Tarnita, C.E., and Nowak, M.A., 2009. Calculating evolutionary dynamics in structured populations. PLoS Comput Biol 5, e1000615.

Newman, M.E.J., Strogatz, S.H., and Watts, D.J., 2001. Random graphs with arbitrary degree distributions and their applications. Phys Rev E 64, 026118.

Nowak, M., and Sigmund, K., 1989. Oscillations in the evolution of reciprocity. Journal of Theoretical Biology 137, 21-26.

Nowak, M.A., 2006a. Evolutionary Dynamics: Exploring the Equations of Life. Harvard University Press, Cambridge, MA.

Nowak, M.A., 2006b. Five Rules for the Evolution of Cooperation. Science 314, 1560-1563.

Nowak, M.A., and Sigmund, K., 2004. Evolutionary Dynamics of Biological Games. Science 303, 793-799.

Nowak, M.A., Tarnita, C.E., and Antal, T., 2010a. Evolutionary dynamics in structured populations. Philos Trans R Soc Lond B Biol Sci 365, 19-30.

Nowak, M.A., Tarnita, C.E., and Wilson, E.O., 2010b. The evolution of eusociality. Nature 466, 1057-62.

Ohtsuki, H., Hauert, C., Lieberman, E., and Nowak, M.A., 2006. A simple rule for the evolution of cooperation on graphs and social networks. Nature 441, 502-505.

Pacheco, J.M., Traulsen, A., and Nowak, M.A., 2006. Active linking in evolutionary games. Journal of Theoretical Biology 243, 437-443.

Paperin, G., Green, D.G., and Sadedin, S., 2011. Dual-phase evolution in complex adaptive systems. J R Soc Interface 8, 609-629.

Perc, M., and Szolnoki, A., 2010. Coevolutionary games--A mini review. Biosystems 99, 109-125.

Poncela, J., and et al., 2009. Evolutionary game dynamics in a growing structured population. New Journal of Physics 11, 083031.

Poncela, J., Gomez-Gardenes, J., Floria, L.M., Sanchez, A., and Moreno, Y., 2008. Complex cooperative networks from evolutionary preferential attachment. PLoS One 3, e2449.

Rainey, P.B., and Rainey, K., 2003. Evolution of cooperation and conflict in experimental bacterial populations. Nature 425, 72-4.

Rozenberg, G., 1997. Handbook of graph grammars and computing by graph transformation: volume I: Foundations. World Scientific, River Edge, NJ, USA.

Santos, F.C., Pacheco, J.M., and Lenaerts, T., 2006. Cooperation Prevails When Individuals Adjust Their Social Ties. PLoS Comput Biol 2, e140.

Scheffer, M., Bascompte, J., Brock, W.A., Brovkin, V., Carpenter, S.R., Dakos, V., Held, H., van Nes, E.H., Rietkerk, M., and Sugihara, G., 2009. Early-warning signals for critical transitions. Nature 461, 53-9.

Schweitzer, F., Fagiolo, G., Sornette, D., Vega-Redondo, F., Vespignani, A., and White, D.R., 2009. Economic networks: the new challenges. Science 325, 422-5.

Sole, R.V., Pastor-Satorras, R., Smith, E., and Kepler, T.B., 2002. A model of large-scale proteome evolution. Adv Complex Syst 5.

Sornette, D., 2003. Why stock markets crash : critical events in complex financial systems. Princeton University Press, Princeton, N.J.

Stiglitz, J.E., 2010. Contagion, Liberalization, and the Optimal Structure of Globalization. Journal of Globalization and Development 1, 2.

Szabó, G., and Fáth, G., 2007. Evolutionary games on graphs. Phys Rep 446, 97-216.

Tarnita, C.E., Antal, T., Ohtsuki, H., and Nowak, M.A., 2009a. Evolutionary dynamics in set structured populations. Proc Natl Acad Sci U S A 106, 8601-4.

Tarnita, C.E., Ohtsuki, H., Antal, T., Fu, F., and Nowak, M.A., 2009b. Strategy selection in structured populations. J Theor Biol 259, 570-81.

Traulsen, A., Shoresh, N., and Nowak, M., 2008. Analytical Results for Individual and Group Selection of Any Intensity. Bull Math Biol 70, 1410-1424.

Traulsen, A., Semmann, D., Sommerfeld, R.D., Krambeck, H.J., and Milinski, M., 2010. Human strategy updating in evolutionary games. Proc Natl Acad Sci U S A 107, 2962-6.

Travisano, M., and Velicer, G.J., 2004. Strategies of microbial cheater control. Trends in Microbiology 12, 72-78.

Vazquez, A., Flammini, A., Maritan, A., and Vespignani, A., Modeling of protein interaction networks, 2001.

Wakano, J.Y., Nowak, M.A., and Hauert, C., 2009. Spatial dynamics of ecological public goods. Proceedings of the National Academy of Sciences 106, 7910-7914.

Wakeley, J., 2008. Coalescent Theory: An Introduction. Roberts & Company Publishers, Greenwood Village, Colorado.

Watts, D.J., and Strogatz, S.H., 1998. Collective dynamics of 'small-world' networks. Nature 393, 440-2.